\documentclass{article}

\usepackage{arxiv}
\usepackage{amsmath}

\usepackage[utf8]{inputenc} 
\usepackage[T1]{fontenc}    
\usepackage{hyperref}       
\usepackage{url}            
\usepackage{booktabs}       
\usepackage{amsfonts}       
\usepackage{nicefrac}       
\usepackage{microtype}      
\usepackage{lipsum}
\usepackage{graphicx}
\graphicspath{ {./figs/} }

\usepackage{outlines}
\usepackage{subcaption}
\usepackage{xcolor}
\usepackage{hyperref}
\hypersetup{backref=page}
\usepackage{cleveref}

\usepackage[numbers, sort&compress]{natbib}
\usepackage{xspace}

\newcommand\skygest{\textsc{Paper Skygest}\xspace}
\newcommand{\arxiv}{arXiv\xspace}

\begin{document}

\title{Paper Skygest: Personalized Academic Recommendations on Bluesky}

\author{
 Sophie Greenwood \\
  Cornell Tech\\
  New York, NY\\
  \texttt{sjgreenwood@cs.cornell.edu} \\
   \And
 Nikhil Garg \\
 Cornell Tech\\
  New York, NY\\
  \texttt{ngarg@cornell.edu} \\
}

\date{}
\maketitle

\begin{abstract}
We build, deploy, and evaluate \skygest, a custom personalized social feed for scientific content posted by a user's network on Bluesky and the AT Protocol. We leverage a new capability on emerging decentralized social media platforms: the ability for anyone to build and deploy feeds for other users, to use just as they would a native platform-built feed. To our knowledge, \skygest is the first and largest such continuously deployed personalized social media feed by academics, with over 50,000 weekly uses by over 1,000 daily active users, all organically acquired. First, we quantitatively and qualitatively evaluate \skygest usage, showing that it has sustained usage and satisfies users; we further show adoption of \skygest increases a user's interactions with posts about research, and how interaction rates change as a function of post order. Second, we share our full code and describe our system architecture, to support other academics in building and deploying such feeds sustainably. Third, we overview the potential of custom feeds such as \skygest for studying algorithm designs, building for user agency, and running recommender system experiments with organic users without partnering with a centralized platform.
\end{abstract}

\section{Introduction}
 Social media recommendation feeds -- on Facebook, Twitter/X, Bluesky, Reddit, TikTok, etc -- mediate individual and societal consumption and discussion of news stories, scientific content, and entertainment \citep{flaxman2016filter,shearer2021more,weeks2017online}. These feeds are largely \textit{centrally designed and deployed}, with users given the choice between only a few ``algorithmic'' feeds developed by the central platform and a ``non-algorithmic'' version consisting of posts by accounts the user follows, in reverse chronological order.
 
 This centralization limits academic study of algorithm design and causal effects of feeds on such platforms \citep{lazer2020studying,mosleh2022field,piccardi2024reranking}.
(1) First, due to bureaucratic and engineering constraints: researchers must either (a) partner with the central platform to study how algorithm design affects organic users \citep{saveski2022perspective, kramer2014experimental,huszar2022algorithmic,lin2024reducing,jilke2019using,nyhan2023like,doi:10.1126/science.abp9364,doi:10.1126/science.add8424}, or (b) externally run controlled lab experiments on simulated platforms or develop a custom layer on top of a platform (such as through browser extensions) \citep{cen2024, salganik2006, piccardi2024reranking,piccardi2024socialmediaalgorithmsshape,milli2025engagement,voggenreiter2024role,jia2024embedding,epstein2022yourfeed,jahanbakhsh2025value,trumanplatform,Jagayat2024}. Platform partnership opportunities are rare,\footnote{As \citet{gonzalez2023asymmetric} state, ``There is virtually no research comparing what people
could potentially see within platforms and what they actually see [with the exception of \citet{bakshy2015exposure}].''} and experimental designs are constrained by the platform's interests. Lab experiments require recruiting a set of participants (often paid), and the design space for browser extensions is often limited to \textit{re-ranking} content surfaced by the platform's feeds \citep{piccardi2024socialmediaalgorithmsshape,piccardi2024reranking}.\footnote{As \citet{piccardi2024reranking} explain regarding browser experiments, ``without access to the complete platform inventory, up-ranking content in the feed presents more challenges than down-ranking content. The content returned by the server is obtained by scoring the inventory, which includes the entire candidate set of potential posts available for display. Without the full set or a meaningful approximation, the intervention can operate only on the pre-selected content delivered to the user, representing the top items the curation algorithm has already prefiltered.''} Maintaining continuous usage for external platforms is a further challenge.
(2) Second, and more fundamentally, a common design characteristic is that a single feed algorithm often surfaces \textit{all} types of content -- whether political, research, entertainment, or social. Thus, studying algorithmic amplification -- either with a platform or using a browser extension -- is often limited to topics such as news consumption and polarization, which are assumed to affect a large fraction of the overall user base. 

\begin{figure*}[t]
    \begin{subfigure}{.45\linewidth}
        \centering
        \includegraphics[width=\linewidth]{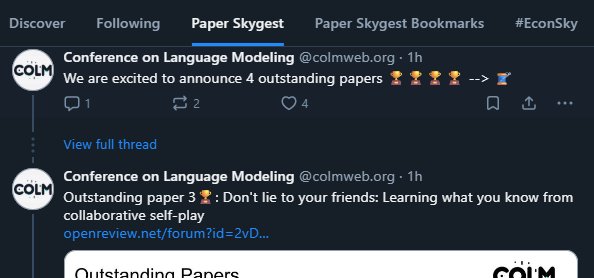}
        \caption{\skygest as it appears to a user in the Bluesky interface.}
        \label{fig:skygestsetup:feedinbluesky}
    \end{subfigure}\hspace{0.7cm}
    \begin{subfigure}{.45\linewidth}
        \centering
        \includegraphics[width=\linewidth]{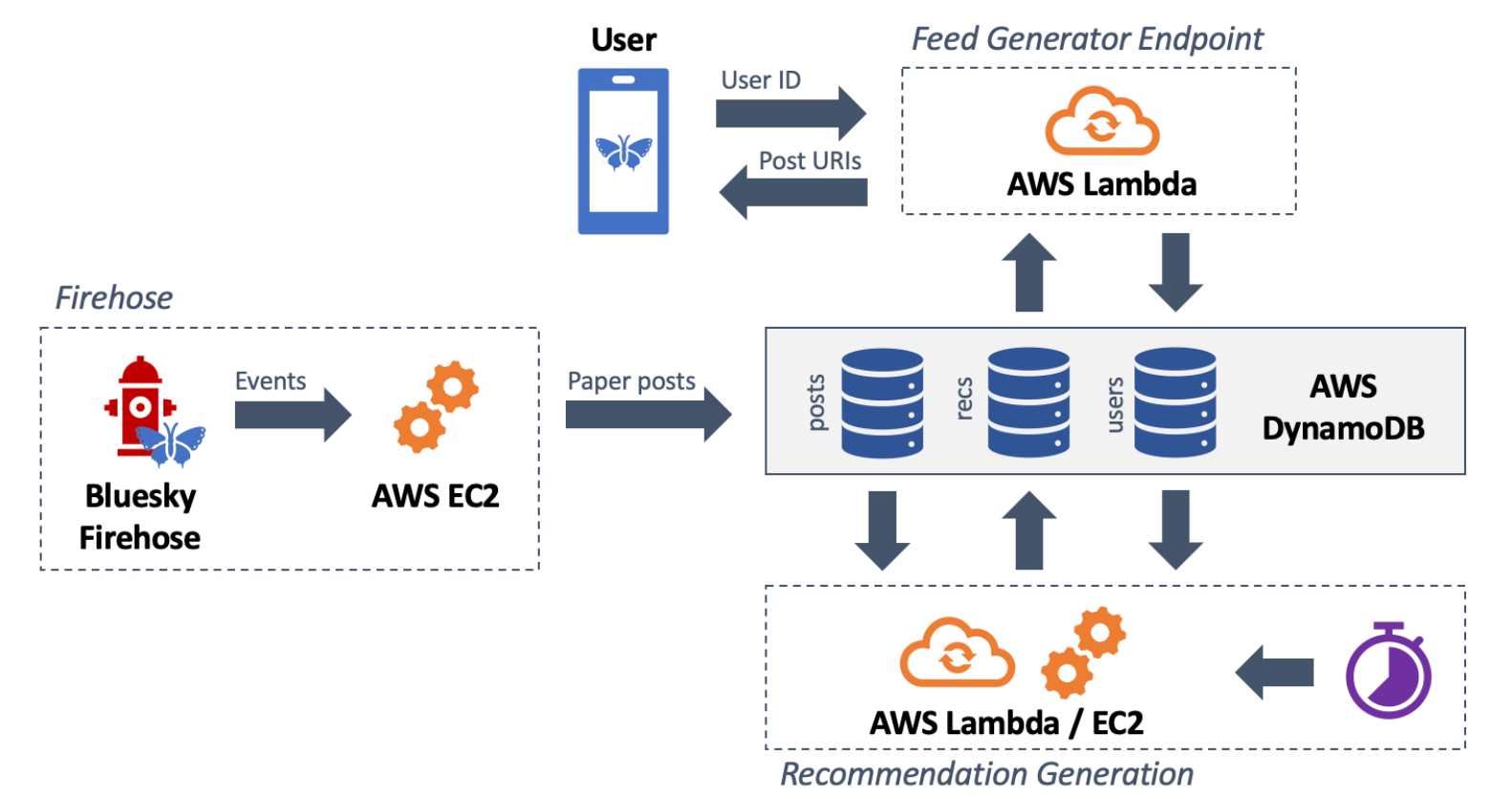}
        \caption{\skygest Infrastructure. }
        \label{fig:skygestsetup:infrastructure}
    \end{subfigure}
    \caption{(a) A user's perspective of using \skygest: the user switches to the feed and sees a list of posts about research papers (including the broader thread) by people in their following network. (b) A diagram of the \skygest backend, which is comprised of three main components: (1) the \textit{feed generator endpoint}, which serves cached recommendations to users, (2) the \textit{firehose}, which listens for and stores paper posts, and (3) the \textit{recommendation generation} module, which precomputes the paper posts to show to each \skygest user at regular intervals and caches them.}
    \label{fig:skygestsetup}
\end{figure*}

This work leverages a new opportunity for social media research that allows academic researchers to flexibly control user recommendations in a real platform environment: \textit{custom feeds} on Bluesky \citep{kleppmann2024}. Bluesky is an emerging social media platform, on the AT Protocol, where users publish and interact with short-form text posts (like Twitter/X), with about 40 million total users and over 1 million daily active users as of October 2025. Bluesky allows anyone to host arbitrary recommendation algorithms, that other users can (organically) choose to use -- fulfilling a desire for such decentralized ``middleware'' \citep{fukuyama2020middleware}. As shown in \Cref{fig:skygestsetup:feedinbluesky}, custom feeds pinned by a user are shown alongside platform-provided feeds. For example, there are popular feeds that are dedicated to trending news, various academic communities (machine learning/AI, Economics), and ``Quiet Posters,'' which prioritizes users who do not post much. Tools such as \url{Graze.social} allow anyone to build feeds without writing code and host thousands of custom feeds, many with tens of thousands of users. 

We leverage this opportunity, and make three contributions. (1) First, we build, deploy, and evaluate \skygest, a \textit{personalized} custom feed for scientific content: it shows posts about research publications by those in the user's following network. To our knowledge, \skygest is the first (and perhaps the only, to date) personalized social media feed deployed at scale for organic usage on a major social media platform either by academics or for scientific content.\footnote{Feeds such as the Science feed or various AI/ML feeds show the same content to everyone. In general, building personalized feeds is much more challenging than non-personalized ones, which can be built using ``no code'' solutions such as Graze.} \skygest serves scientists and others who discuss scientific content on Bluesky. Social media, especially ``academic Twitter,'' has long driven consumption of academic content. Especially since Fall 2024, many academics have migrated from Twitter/X to Bluesky \citep{shiffman2025}, although adoption is heterogeneous between research communities \citep{quelle2025leaving}. Altmetric, which tracks social media mentions of published papers, estimated that for most of March 2025 Bluesky hosted as many or more daily links to published academic research as did Twitter/X \citep{altmetricblog}.

\skygest has continuously run since its public launch in March 2025 and has garnered substantial organic user adoption, with over 50,000 weekly uses by over 1,000 daily active users, totaling over 2 million uses to date. We show qualitatively and quantitatively that users find \skygest useful and continuously return to the feed; we further use \skygest data to show how users consume and interact with scientific content, and that using \skygest shifts consumption patterns toward scientific content.

(2) To enable other academics to build, deploy, and evaluate such feeds for themselves, we describe the engineering architecture behind \skygest. We describe engineering lessons learned, especially for other academics aiming to build such feeds for research. Our code is available at 
\url{https://github.com/Skygest/PaperSkygest}.

(3) We provide a vision for how academics can use new capabilities provided by the AT Protocol and Bluesky to run large-scale randomized field experiments on algorithmic design with organic users, without partnering with a platform. This new capability enables deployments and evaluations of new approaches to support user agency, discovery of niche topics, and feed ranking beyond engagement optimization. 

This paper is organized as follows. Related work is in \Cref{sec:related}. In \Cref{sec:skygestusage}, we overview \skygest and present quantitative usage metrics and qualitative feedback; we further leverage \skygest data to show how users view, consume, and share research on Bluesky. \Cref{sec:architecture} contains a description of our system architecture. In \Cref{sec:vision}, we describe how the AT Protocol, and \skygest in particular, enables new forms of social media algorithm design and evaluations. We conclude in \Cref{sec:conclusion}. 

\begin{figure*}[t!]
    \begin{subfigure}{.4\linewidth}
        \centering
        \includegraphics[width=\linewidth]{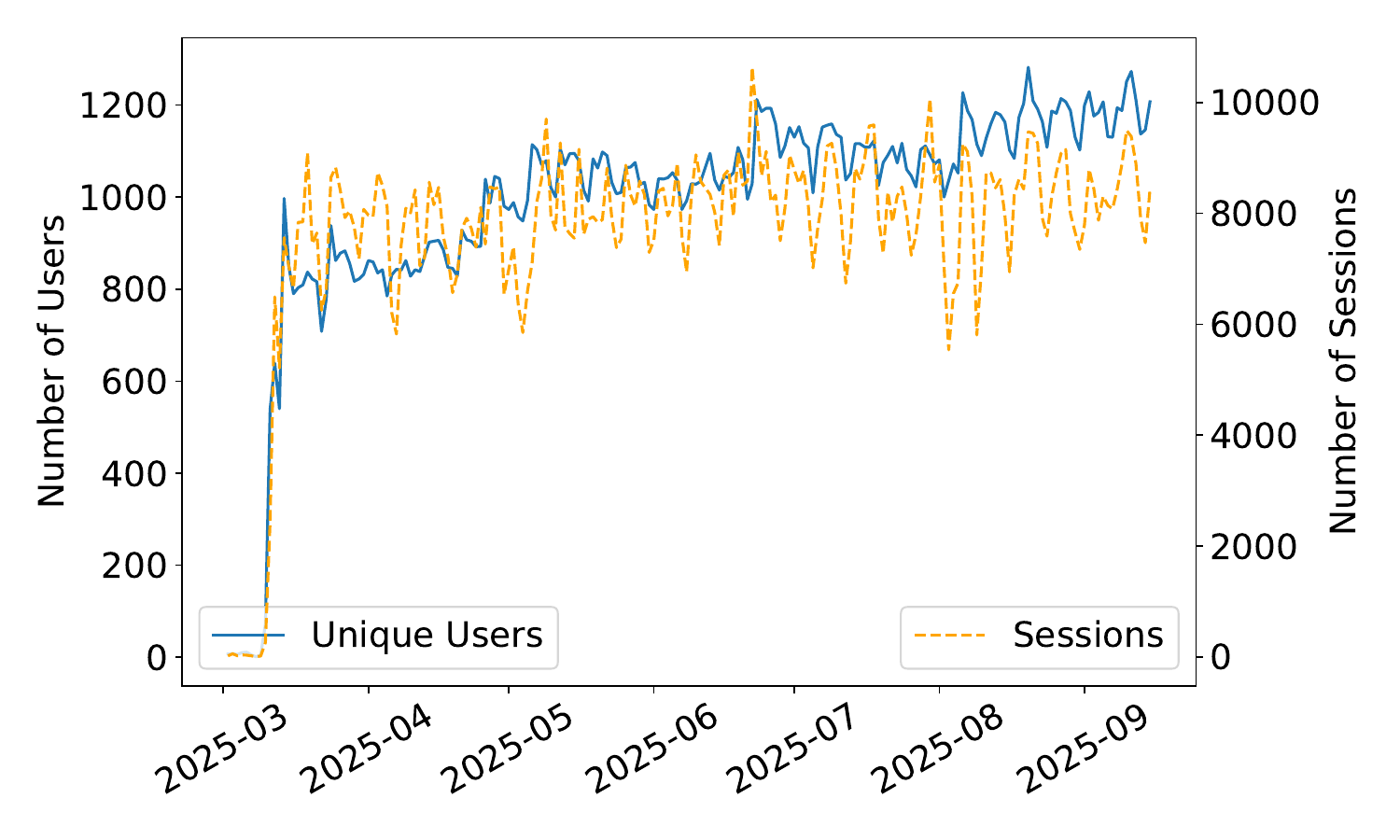}
        \caption{Daily active users and accesses over time.}
        \label{fig:usages:users}
\end{subfigure}\hfill
\begin{subfigure}{.55\linewidth}
    \centering
    \includegraphics[width=\linewidth]{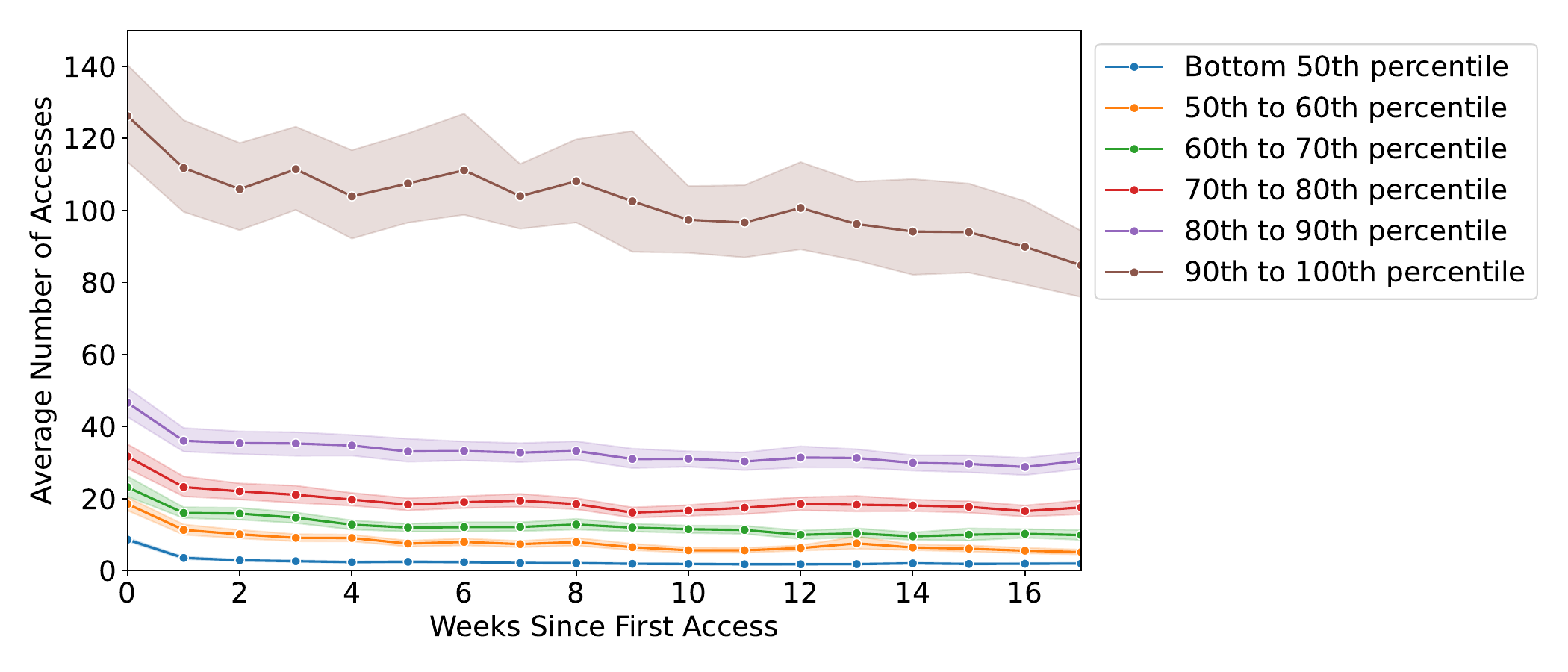}
    \caption{Individual usage persistence over time}
    \label{fig:stickiness}
\end{subfigure}
\caption{\skygest usage data over time. (a) Figure \ref{fig:usages:users} shows, for each day since launching, the number of unique accounts that used \skygest and the number of sessions on \skygest (where a session is defined as a request from Bluesky for the first page of posts). The number of unique users has increased over time, while the daily number of sessions has remained stable. Both user and session counts vary throughout the week, with high counts on weekdays and low counts on weekends. (b) Figure \ref{fig:stickiness} shows average usage trajectories for users over time, where users are grouped by their percentile of total \skygest accesses. We consider usage data between May 21, 2025 and September 16, 2025, for users who accessed \skygest for the first time prior to May 21, 2025; we average across users and report $95\%$ confidence intervals.}
\end{figure*}

\section{Related Work}
\label{sec:related}
\paragraph{Deployed Recommenders from Academia} Industry building and deploying recommender systems is common, with many publications on experimental evaluations thereof and engineering lessons learned \citep{cai2023two,wu2025graphhash,10.1145/3543507.3583202,10.1145/3589335.3651486}. In contrast, long-term recommender deployments with organic users are rare coming out of academia. Engineering and maintenance costs, the difficulty of attracting organic usage, and academic incentives are key barriers. A notable exception is MovieLens and other deployed recommenders from the GroupLens academic group \citep{konstan1997grouplens,10.1145/371920.372071,10.1145/2827872,10.1145/2792838.2800179,10.1145/2792838.2800195,panciera2010lurking}; more recent examples include ScholarInbox \citep{flicke2025scholar} and POPROX \citep{burke2025conducting,higley2025news}. Most related to our work is ScholarInbox, who build a self-standing platform to recommend research papers, primarily to computer scientists; \citet{flicke2025scholar} describe their system design and survey their users. To our knowledge, \skygest is the first continuously deployed personalized social media feed from academia. To support others to do the same, we contribute an example system design and open source code.

\paragraph{Social Media Feed and Recommender Design}
There is substantial academic interest in designing algorithmic feeds \citep{piccardi2024reranking} -- especially those that go beyond engagement optimization in supporting user agency and societal values \citep{malki2025bonsai,cunningham2025ranking,milli2023choosing,MilliOptimizing2021,jahanbakhsh2025value,kleinberg2024challenge,weyl2025prosocial,stray2021you,stray2024building}.
 However, as \citet{piccardi2024reranking} state, ``these questions have
remained largely unanswerable to the group of people without personal access to hundreds of millions of users.'' More generally, algorithm design for recommenders, including for for academic content, is of substantial interest \citep{10.1145/3491102.3517470,greenwood2024,peng2024reconciling,patro2022fair, rectostratusers, kleinberg2024calibratedrecommendationsusersdecaying, piccardi2024socialmediaalgorithmsshape}.
However, as \citet{higley2025news} discuss, there is a substantial gap between conducting recommender system research and building one. As discussed above, \skygest provides a deployment avenue for such approaches, and more broadly, we aim to help close the gap for academic research and deployments for feed design. 

Recent work has also leveraged the opportunities afforded by the decentralized nature of emerging social media \citep{fukuyama2020middleware,oshinowo2025seeing,hwang2025governing}, primarily by prototyping and evaluating with a small set of (potentially paid) subjects on Bluesky and Mastodon \citep{malki2025bonsai,zhang2025understanding}. However, to our knowledge, these systems have not been deployed at scale for organic usage overtime. 

\paragraph{Empirical Analyses of Bluesky and Scientific Discovery on Social Media} A number of recent studies have performed observational analyses of the Bluesky ecosystem \citep{balduf2024, failla2024, jeong2024, smith2025bluestartlargescalepairwise,quelle2025leaving}, leveraging its fully open (and monetarily free) data ecosystem. For example, \citet{quelle2025customfeeds} analyze the large ecosystem of custom feeds. 
A particular focus has been analyzing how academics have adopted the platform \citep{quelle2025leaving,zheng2025science, ARROYOMACHADO2025101700}; for example, \citet{quelle2025leaving} identifies over 300,000 academic Bluesky users. 
More generally, academic behavior on social media -- and its implications on science -- has been a substantial area of study \citep{alabrese2024politicized,garg2025political,mongeon2023open,guenther2023science,darling2013role,su2017information,junger2020does,ke2017systematic,van2011scientists,collins2016scientists,bik2013introduction}, studying how social media usage and visibility affects scientific outcomes and academic productivity. One notable example is \citet{qiu2024social}, who experimentally evaluated how social media affects job market outcomes, by assigning EconTwitter influencers to randomly retweet posts about job market papers -- women in the treatment group, on average, received 0.9 additional job offers;  similarly, \citet{branch2024controlled} experimentally find that papers promoted by leading science communicators were downloaded about 3$\times$ more after tweeting and had 81\% higher Altmetric scores, but the increase in citations three years later was not statistically significant.

To this literature, we contribute behavioral observations about the effects of social media rankings; for example, a user is 4$\times$ more likely to ``like'' a post that appears first in a their \skygest feed than one that appeared, at highest, fifth. As we discuss, \skygest further provides a deployment setting to experimentally measure such effects, such as through running randomized controlled trials on algorithm design.

\section{Paper Skygest and Usage}
\label{sec:skygestusage}

\begin{outline}
    In this section, we give further details on Bluesky's custom feeds, and  describe \skygest. We then provide statistics on our feed's usage, and show qualitative feedback from Skygest users along with usage statistics. Finally, we illustrate behavioral effects in how users interact with posts and how adopting \skygest corresponded with a shift in behavior.
\end{outline}

\subsection{Paper Skygest Overview}

Custom feeds on Bluesky are recommendation algorithms which are designed and hosted by people external to the Bluesky organization. These feeds are not simply a \textit{feature} Bluesky offers -- users are \textit{encouraged} to pin custom feeds as they are a core component of Bluesky's ecosystem, offering users ``algorithmic choice'' \citep{kleppmann2024}. When a user discovers a feed she likes -- through either her social network or Bluesky's feed search tool -- she can pin it to the feed navigation bar shown in Figure \ref{fig:skygestsetup:feedinbluesky}. While using the app, the user can swipe left and right to switch between pinned feeds, and can reorder the feeds in the navigation bar as she chooses. 

Under the hood, each time a user navigates to a custom feed such as \skygest in the navigation bar, Bluesky sends the user's identifier to the custom feed's \textit{feed generator}, and the feed generator returns a ranked list of posts to show the user. The feed generator's only responsibility is to determine the list of posts to be shown, and Bluesky places no constraints on this list. 

\skygest currently shows a user posts about academic papers from accounts she is following, in reverse-chronological order. \skygest includes a post if its text or metadata include a link to a site hosting preprints or published manuscripts, or contain at least three keywords related to academic paper announcements -- we make this logic open-source through GitHub. A more detailed system architecture is provided in \Cref{sec:architecture}, and we provide a schematic diagram in Figure \ref{fig:skygestsetup:infrastructure}.

We launched \skygest publicly on March 10, 2025. Between then and September 17, 2025, an average of 1,029 unique users visited \skygest each day, and an average of 1,614 users visited \skygest each week, with growing numbers of active users over time; in total, \skygest was used 1,526,806 times in the time period. To launch \skygest, we simply posted an announcement on Bluesky and encouraged users in our network to spread the word and repost our announcement. Our user base then grew naturally over time as users reposted or recommended \skygest to other users. Notably, we did not take out advertisements for \skygest or recruit paid participants -- our users are organic.

\paragraph{Consent and Research Ethics Approval} This research was approved by Cornell University IRB under Protocol \#0149634. For a user's first five visits, \skygest shows a post thread explaining that we are conducting research, including information about data collection and consent.\footnote{This consent information was added several months after initial deployment. When it was added, all current users were shown the consent thread for their next five visits.} Users may request their data to be removed and excluded at any time while still maintaining feed access. In this work, we include a user in our analysis if she completes these five visits and does not opt out of data collection. We exclude our research team from analysis. In all analyses and usage statistics, we restrict to data prior to September 17, 2025. 

\subsection{Quantitative Usage Metrics}

\paragraph{\skygest has substantial organic usage} In Figure \ref{fig:usages:users}, we plot usage metrics for \skygest since its launch in March 2025; our user base has increased over time from an average of 889 users per day in April, to an average of 1,147 users per day during August. \skygest has maintained a high level of traffic; on average \skygest is checked 7,994 times per day, for a total of 1,526,806 sessions  during the studied time period. If we exclude users with outlying access patterns (over 20,000 daily accesses each), \skygest is accessed 5,622 times on average per day for a total of 1,073,812 sessions. In all subsequent analyses, we restrict to the 3,241 users with more than 5 but fewer than 20,000 total accesses each.\footnote{Bluesky sometimes pre-loads the feed even when the user has not viewed the feed, which would not be considered a real access. With the data we currently have access to, we cannot observe the difference between a true access and a pre-load. Therefore, these numbers are potentially overestimates of true usage.}

\paragraph{\skygest users return over time} \skygest also has many users who have not ``churned'' off the feed. Each week, an average of 810 users use \skygest for at least five days -- a large cohort are dedicated users of the feed. In Figure \ref{fig:stickiness}, we show user trajectories in the 16 weeks after the first time they accessed \skygest, with different cohorts defined by how much they have used \skygest overall. As illustrated in the bottom Blue line, about half of \skygest users (who make it past the 5 consent posts) approximately do not return to the feed after the first week. However, for the remainder groups, average usage 16 weeks after the first access is only slightly lower than average usage in the second week; thus, while the frequency of access differs substantially across users, individual users tend to return to \skygest over time.

\subsection{Qualitative Feedback}
Since the launch of our feed, we have received substantial interest and positive feedback; almost all user growth stems from word of mouth by \skygest users -- we do not conduct online advertising, and to our knowledge, feed discovery avenues on Bluesky are not a big channel for adoption. Here, we analyze qualitative feedback, by users posting about \skygest on Bluesky. The below quotes are organically generated public posts on Bluesky by users independent of the research team, who posted without being asked. We received permission from each quoted user to include their post; we include their handle and link to post if given explicit permission to do so, and otherwise default to quoting the post anonymously. Emojis are represented by \texttt{:emoji-name:}. 

Many users emphasize the usefulness of \skygest in highlighting posts about research papers that might otherwise be missed or buried under other content:
\begin{itemize}
    \item \textit{This might be the most useful thing I have come across in social media - a personalized feed of academic papers filtered by your follower network! Highly recommend. \#academicsky} -- \href{https://bsky.app/profile/suryaganguli.bsky.social/post/3ll3tz3yaws2r}{@suryaganguli.bsky.social}
    \item \textit{This feed really does make bsky useful again for following research. (My home had descended into full politics for the last few weeks.)} -- \href{https://bsky.app/profile/nsaphra.bsky.social/post/3lkcwiobubc2r}{@nsaphra.bsky.social}
        \item \textit{This is the first feed that has felt like proper academic twitter (love it or hate it)} -- \href{https://bsky.app/profile/eugenevinitsky.bsky.social/post/3lk4qyibags2s}{@eugenevinitsky.bsky.social}

    \item \textit{This has potential for academics who miss old Econtwitter.} -- anonymous
    \item \textit{This is awesome ! \texttt{:thumbs-up:} One scroll of the feed and I am seeing several papers and posts that I missed via my Discovery feed.} -- anonymous
\end{itemize}

Several users noted in particular that \skygest illustrates the power of externally developed feeds, as benefiting both users and researchers. For example one user, 
in response to our posts disclosing that we are conducting research, said

\begin{quote}
   \textit{I have been thinking for a while about how AT Proto and Bluesky can benefit researchers[.] One way is to help surface relevant papers, which is exactly what Skygest is doing[.] But AT Proto/Bluesky can in itself be used to conduct research, which is also what Skygest is doing!} -- \href{https://bsky.app/profile/o.simardcasanova.net/post/3lppclylnj22g}{@o.simardcasanova.net}
\end{quote}
Another user said
\begin{quote}
\textit{This is \texttt{:sparkles:}excellent\texttt{:sparkles:} as it picks up all the papers that folks mention to each other in replies, as well as those in top level posts. Brilliant way to make use of the power of feeds.} -- anonymous
\end{quote}

\subsection{User Behavior}

\begin{figure}[t]
        \centering
        \includegraphics[width=0.8\linewidth]{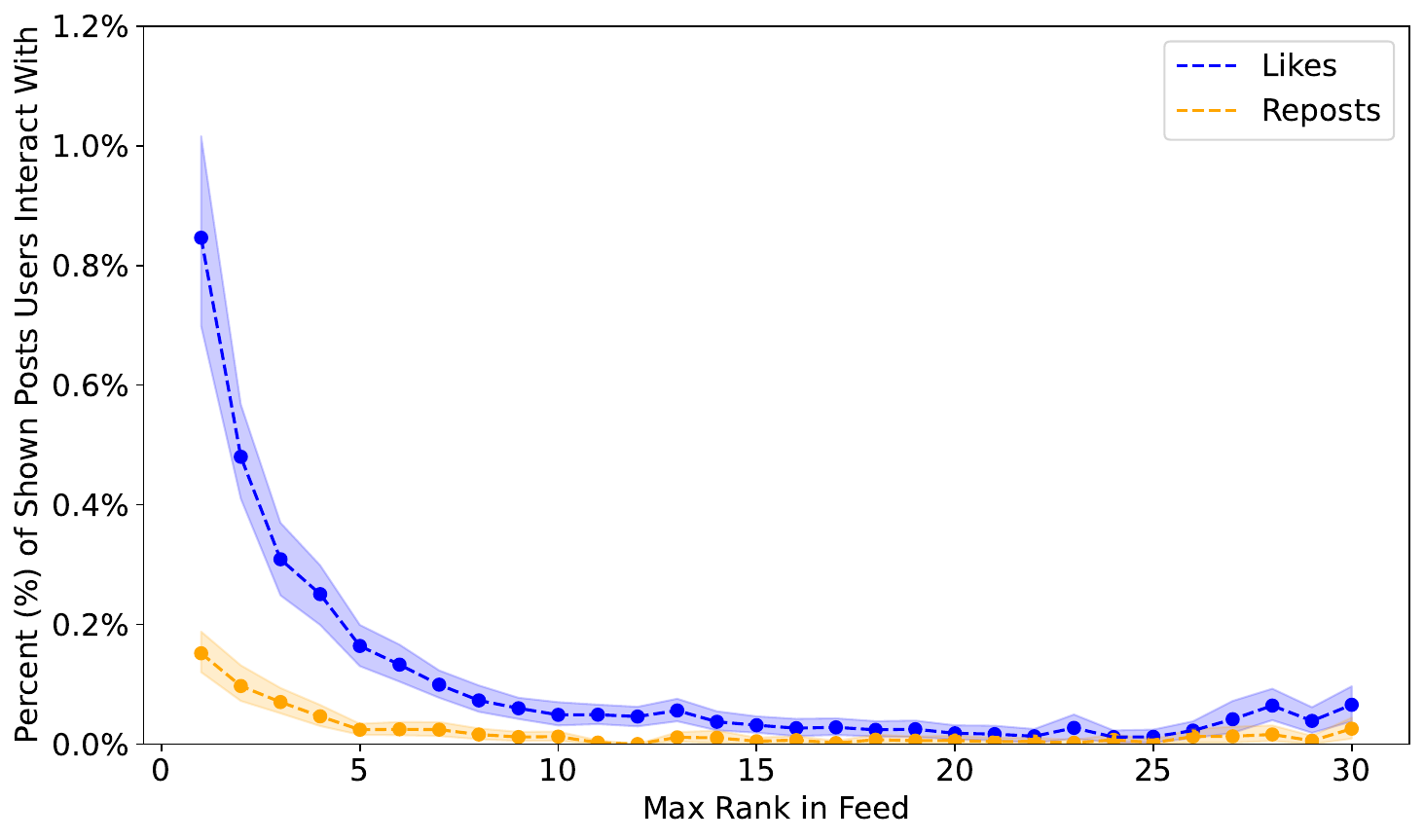}
        \caption{The rate at which users like or repost a post, as a function of how high it appeared in their feed when they switched to it (since users may access the feed multiple times with overlap in posts, we use the highest position that the post appeared). There are strong feed positioning effects. We consider only posts, likes, reposts, and accesses up to September 16, 2025, and filter to user likes and reposts that occurred within 30 seconds of the user accessing \skygest. Bluesky typically requests pages of size 10 and 30; we use access data for cases where Bluesky requests 30 posts. This results in 5,201 likes and 1,046 reposts. We take the average across post-user pairs for each rank, and obtain clustered $95\%$ confidence intervals by generating 1,000 bootstrap samples of users. The same pattern holds when we increase the interval considered above 30 seconds (see Appendix \ref{app:funnel}).}
        \label{fig:funnel}
\end{figure}

 We now leverage data on \skygest accesses and our corpus of posts about academic papers to study how scientific content is shown to and interacted with by \skygest users. These are examples of the analyses that \skygest facilitates -- and custom feeds on Bluesky more generally -- even before running randomized experiments. We refer to the corpus of posts that pass our filter as ``paper posts.''

The engagement data used below is streamed from the public firehose as we describe in Section \ref{sec:architecture}, which provides timestamped records of each instance a user interacts with a paper post. To obtain records of likes of general (non-paper related) Bluesky posts, we query public PDS data through Bluesky's API.\footnote{The public records do not include information about whether the user was using \skygest at the time of the interaction. In October 2025, Bluesky began to send private records to feed designers indicating when a user interacted with a post while using the designer's feed. We do not use this data in our analyses.}

\paragraph{Engagement as a function of feed ranking} An oft-observed phenomenon in recommender systems and feed design is that item ordering matters -- users are more likely to interact with items shown first. For example, \citet{arxivpositionaleffects,haque2010last} find in 2009 in several physics categories on \arxiv that papers at the top of the daily digest email have over 50\% more citations and visibility, with similar positive effects of being last (compared to being in the middle). Similarly, \citet{nber2017} found that being listed first in the NBER\footnote{A prestigious economics organization that sends out working papers by its members} email of working papers leads to a 33\% increase in views and 27\% increase in citations, even though papers are ordered via a process that the email generators view as random.

To our knowledge, social media companies have not generally released individual-post-and-user level view and interaction data, and so it is unknown how strong these effects are for scientific content on social media. \Cref{fig:funnel} illustrates how users' rate of interacting with a post changes with the maximum rank the post had when shown to the user in \skygest. We observe a sharp drop-off in likes as the maximum rank increases; while there are far fewer reposts, we observe a similar decrease here. In particular, users are much more likely to interact with a post first in their feed -- for example, 4$\times$ more likely to ``like'' a post that appeared first in a their \skygest feed than one that appeared, at highest, fifth. The relative differences are even stronger for reposts. These relationships are much higher than those found in the \arxiv and NBER settings, underscoring the importance of social media algorithm design in mediating scientific consumption. We note that this interaction difference is not directly due to our algorithm prioritizing content that we believe the user will like more, since \skygest currently orders posts in reverse chronological order.\footnote{There may be other differences explaining the effect, such as that feed ordering and feed usage timing correlates jointly with other reasons a user interacts with posts; for example, that they open our feed immediately after a friend posts a paper thread, to interact with that post. Such confounding can only be mitigated through a randomized controlled trial on feed order; we discuss such experiments in \Cref{sec:vision}.}

\paragraph{User engagement with scientific posts after adopting \skygest} 
Next we ask: ``does using \skygest correspond to a change in behavior on Bluesky?'' We might expect that use of \skygest leads to an increase in engagement with academic content, compared to before the user started using \skygest -- since \skygest makes discovery of such posts easier.

To evaluate this, we compare how each user interacted with posts in the windows of time one week before to one week after the user's first time accessing \skygest. We query the Bluesky API for all ``like'' events for \skygest users across all Bluesky posts; we consider users who are already actively engaged with academic content on Bluesky before this window (to mitigate capturing naive effects such as a user first using \skygest immediately after they started using Bluesky). In particular, we consider users who had at least one like of a paper post and one like of a non-paper post in the week before the window (i.e., two weeks to 1 week before they first used \skygest)
For technical data collection reasons and to remove potential bots, we further remove users who had more than 10,000 total likes on Bluesky. After these filters, the following analysis corresponds to 796 (25\%) \skygest users. (We find directionally similar findings if we make the minimum activity threshold larger, though this removes more users). 

For each user, we count the number of likes of paper posts in the week before and the week after first accessing \skygest, $\ell_{\text{paper}}^{\text{before}}(u)$ and  $\ell_{\text{paper}}^{\text{after}}(u)$ respectively. We first consider the difference in likes of paper posts in this window after accessing \skygest: the average difference $\ell_{\text{paper}}^{\text{after}}(u)-\ell_{\text{paper}}^{\text{before}}(u)$ is $0.56$ with standard error $0.12$, yielding a $95\%$ confidence interval of $(0.33, 0.78)$. In other words, on average, a user liked $0.5$ more posts about research papers in the week after adopting \skygest than in the week before, and the difference is statistically significant -- note that this difference occurs even though ``like'' interactions are sparse, with users on average liking less than 1\% of posts about papers that are shown to them.

One limitation of the above analysis is -- even though we filter for users already active on Bluesky -- that user activity may be changing over time, and adoption of \skygest may still correlate with an intention to be more active on Bluesky. Thus, to better capture the effect of \skygest, we compare this increase in interactions with the corresponding change in interactions with posts that are \textit{not} about research; if user activity levels are driving the above effect, then we might expect that interactions with non-research posts \textit{also} increase a corresponding amount. Thus, we next analyze whether the \textit{fraction} of a user's engagement that is with posts about research changes in the week after first using \skygest. Formally, we collect each user's total likes in the window before and after, $\ell_{\text{total}}^{\text{before}}$ and $\ell_{\text{total}}^{\text{after}}$, and compute the difference in the \textit{proportion} of the user's likes that are of paper posts \[\frac{\ell_{\text{paper}}^{\text{after}}(u)}{\ell_{\text{total}}^{\text{after}}(u)} - \frac{\ell_{\text{paper}}^{\text{before}}(u)}{ \ell_{\text{total}}^{\text{before}}(u)}.\] The average difference of proportions is $0.035$ with standard error of $0.0078$, yielding a $95\%$ confidence interval of $(0.019, 0.050)$. In other words, users do indeed shift their interaction toward posts about research after adopting \skygest, and the effect is small but statistically significant (on average, the percentage of their interactions that are on posts about papers increases by 3.5 percentage points).

\paragraph{Content Distributions}
Finally, we compare the academic content shown to users and interacted with on \skygest, compared to the universe of academic content on Bluesky. We are interested in two questions: (1) do \skygest users come from across scientific disciplines (as opposed to, for example, just computer science); and (2) are there differences across (sub-)disciplines in how users interact with shown content? 

First, we find that \skygest users are shown a variety of academic content, reflecting a diverse user base (since content is only shown if a user follows the posting account). We find that $8.9\%$ of posts shown to \skygest users include a link to an article from the general interest journal \textit{Nature} (or a \textit{Nature} subjournal), around $5.8\%$ link to an economics journal or preprint server, and around $14.6\%$ link to a preprint on \arxiv. There is also substantial content from journals belonging to various engineering and humanities disciplines.

\begin{figure*}
    \centering
    \includegraphics[width=\linewidth]{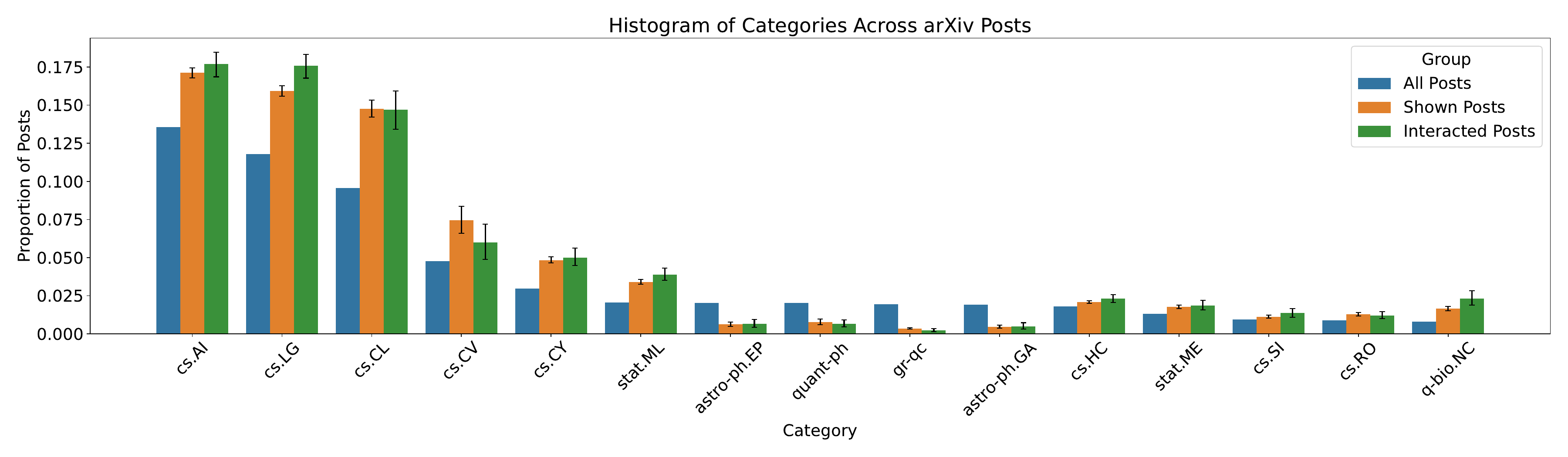}
    \caption{The distribution of \arxiv categories among (a) posts on Bluesky containing \arxiv links, (b) all posts containing \arxiv links shown to \skygest users (allowing duplicate posts across users), and (c) posts containing \arxiv links liked by \skygest users (again allowing duplicates across users). We show categories that are in the top 10 most common categories for at least one of these distributions. We obtain clustered $95\%$ confidence intervals by generating 1,000 bootstrap samples of users. We limit our consideration to posts from non-bot accounts. We identify bots as accounts which mention \texttt{arxiv} or \texttt{bot} in their handle, or have very high posting activity on Bluesky. We use public data from \arxiv \citep{arxiv_org_submitters_2024} to find a post's categories given its \arxiv identifier. There is heterogeneity in the rate at which posts are interacted with, across categories. Note that bars do not add up to one due to ommitted \arxiv categories.}
    \label{fig:arxivcats}
\end{figure*}

Second, we focus on posts that link to \arxiv to study if there are behavioral differences across (sub-)disciplines (since we can retrieve the \arxiv sub-category for each linked paper). For this analysis, we further filter out (do not include) \arxiv posts in our database that are visibly from bot accounts (e.g., accounts that make posts about each new \arxiv paper in a particular category); these bot posts correspond to around $96\%$ of posts with links to \arxiv papers. Of the remaining posts, over $39\%$ are shown to \skygest users at least once. \Cref{fig:arxivcats} compares the distribution of \arxiv categories across posts in the corpus with the distribution of \arxiv categories across posts shown to \skygest users, and again with the distribution of \arxiv categories across user likes of academic posts. Across the board, there is high representation from computer science and specifically artificial intelligence (AI) and machine learning (ML). Furthermore, posts with papers in AI/ML receive a disproportionate number of likes as well as exposure on \skygest compared to the number of posts on Bluesky in the category. Similarly, posts in the cs.CY category (``Computers and Society'') are both disproportionately shown to \skygest users and engaged with upon being shown. On the other hand, posts about quantum physics (quant-ph and gr-gc) are interacted with at lower rates. 

\section{Paper Skygest Architecture}
\label{sec:architecture}

In this section, we describe the backend infrastructure of \skygest, to support others in developing their own feeds. Our code is available at \url{https://github.com/Skygest/PaperSkygest}. We illustrate our architecture in Figure \ref{fig:skygestsetup:infrastructure} -- primarily built on AWS services. As one measure of implementation challenges others might expect: an initial prototype of the feed, based on a prior template \citep{marshalXrepo},\footnote{Our initial code was adapted from an open-source Python template from MarshalX \citep{marshalXrepo}, which provides a self-standing single server feed implementation, with a Peewee SQLite database. We decomposed the system into several interdependent modules in AWS EC2 and Lambda and extended it, for auto-scaling, efficiency, and robustness.} was built in less than two days in December 2024; however, it had multiple-second latency, and only worked for a few users at a time. Our current architecture, as provided, has around 50 millisecond latency, autoscales with users, and supports multiple algorithms -- and was developed over several months and had several test users, before its public launch in March 2025. 

\paragraph{Bluesky Custom Feed Generator Overview} A simplified overview of how Bluesky integrates with custom feed generators is as follows; for further technical details, see Bluesky's documentation \citep{blueskyfeeddocs}. When a user selects a feed to view (e.g., \skygest), Bluesky\footnote{In general, custom feeds can be queried by other clients on the AT Protocol; we refer to Bluesky here for simplicity.} looks up the feed generator (e.g., our system) associated with that feed, and sends a request to that feed generator along with an authentication token which can be used to identify the user. The feed generator determines which posts to show that user, and returns an ordered list of post identifiers -- any post on the AT Protocol can be returned, whether or not the user follows the account associated with the post. Bluesky then shows the user (most of) the posts specified in the returned list; it handles all visualization aspects such as hydration of interaction data; it further handles threading, moderation, and post deletions, and so the list of posts presented may not exactly be the list of posts returned by the generator.  These steps are hidden from the user -- from her perspective, she simply clicks on the feed, and sees a ranked list of posts, all within the Bluesky app.

There are three main components to building a custom feed: (1) handling incoming requests from Bluesky by returning a set of posts, (2) maintaining a corpus of candidate posts to recommend, and (3) implementing recommendation logic determining which posts to show for each user. We give details on how we addressed each of these components for \skygest below. 

\subsection{Feed Generator Endpoint}

A feed generator is a server that hosts a feed; the main requirement of a feed generator is to implement the endpoint \path{/xrpc/app.bsky.feed.getFeedSkeleton}, which returns a list of posts to show. For \skygest, determining which posts to show has high latency -- 75\% of our users' recommendations take over six seconds to generate -- since we need to get a user's follows from the Bluesky API. This is a prohibitively long time for users to wait, so instead we generate recommendations offline and store them in a database, which we then query in real time. Beyond the efficiency gains, decoupling the generation and serving of recommendations also improves robustness -- if an error occurs during recommendation generation, the only consequence on the user's experience is that the user will see stale recommendations. 

Given this, \skygest's implementation of the \path{getFeedSkeleton} endpoint is as follows: First, we resolve the request's authentication token to obtain the user's identifier. If it is the user's first time, we trigger an event to generate the user's recommendations offline, and immediately return a single post prompting the user to refresh after a few seconds. Otherwise, if \skygest has seen the user before, we query an AWS DynamoDB database (``\texttt{recs}'') to get a list of cached recommendations for the user (generated as described below in \Cref{sec:generation}).
Before returning posts, we trigger an event to asynchronously regenerate the user's recommendations, so that if the cached recommendations are stale, the user can refresh after a few seconds and see new recommendations. We also trigger an postprocessing event to asynchronously save the user to a \texttt{users} database, and log the user's access, including the posts shown. For the first five visits by a user, we further show a consent thread, explaining that we are conducting research and how to opt out. This approach balances low real-time latency with delivering fresh content.

\paragraph{Pagination} The \texttt{getFeedSkeleton} request comes with two parameters: \texttt{limit} (the number of posts to return in one page of content) and, if the user has already consumed a page of content, a  \texttt{cursor} (the current page number). We return the posts from index \texttt{cursor} to $\texttt{cursor} + \texttt{limit} - 1$ from the list of posts queried from the \texttt{recs} database, along with $\texttt{cursor} + \texttt{limit}$ as the new cursor.

\paragraph{Cloud Hosting Details} We host this logic on AWS Lambda, a serverless compute service which scales allocated compute resources with usage. This allows us to adapt and scale to different system loads across time in a robust and cost-effective manner. We set up the endpoint on AWS API Gateway, and attach it to the Lambda function. The postprocessing occurs in a separate Lambda function. The recommendation generation and postprocessing events are triggered using AWS SQS. 
If the Lambda function is not called for a few minutes, it will shut down and restart the next time it is invoked; since restarting introduces a delay in serving recommendations, we set up an AWS EventBridge schedule to invoke the function every four minutes to keep the Lambda function active.

In addition to the \texttt{getFeedSkeleton} endpoint, we also publish a feed record with a unique identifier, and implement two other required endpoints (\path{/xrpc/app.bsky.feed.describeFeedGenerator} and \path{/.well-known/did.json}). We again implement these endpoints using AWS Lambda and API Gateway; the code for these two endpoints does not differ significantly from the template \citep{marshalXrepo}.

\subsection{Firehose} To assemble a corpus of Bluesky posts, we set up an AWS EC2 server to subscribe to the Bluesky firehose. Put simply, the firehose is a stream of public ``events'' on Bluesky, such as a user creating or liking a post, or following another user \citep{kleppmann2024}. In particular, we listen for post creation events. We examine each new post's text and metadata (including embedded links), and filter to posts that are about an academic paper -- namely, posts that either (a) contain a link to an academic journal or platforms which host preprints and working papers, or (b) have at least three keywords relating to announcing new academic work. We also make this filter open-source. We store these posts in an AWS DynamoDB database (``\texttt{posts}''), and maintain an index of post authors sorted by creation date.

In addition to post creations, we also listen for several other events on the firehose, including user interactions (for example, a user likes a post). We maintain a local database on EC2 of the unique identifiers of all posts about papers in our dataset, and filter to interactions with paper posts. We save these filtered interactions to additional DynamoDB databases. We also listen for post deletions to mark these posts as deleted in our posts database.

For efficiency, after ingesting relevant events, we add the event to a queue. We set up parallel worker processes to take events from the queue to filter and write to the database. 

\subsection{Recommendation Generation}
\label{sec:generation}
The recommendation generation process is what is used to calculate recommendations for each user, given the database of paper posts and a set of recommendation algorithms. It is decomposed into two components: the dispatch module, which assigns \skygest users to batches, and the generation module, which generates recommendations for each batch in parallel.

\paragraph{Recommendation Dispatch Module} The recommendation dispatch module queries the \texttt{users} database for a list of \skygest users. The module randomly assigns these users into batches of 20, and triggers an asynchronous invocation of the generation module for each batch. This module is scheduled to run every 20 minutes.

\paragraph{Recommendation Generation Module} 

For each user in a batch, this module generates recommendations and saves them to the \texttt{recs} database, the database queried by the feed generator queries when a user requests the feed. 

For each user, the generation module runs one or more recommendation algorithms. In the default \skygest experience, users are shown recommendations from the ``reverse chronological'' algorithm, which simply shows posts from accounts the user follows in reverse chronological order. Given a user's identifier, we request the accounts that user is following from Bluesky's public API. Then, for each of these accounts, we query the author index of \texttt{posts} to find the ten most recent posts from that account. We then sort all of these posts chronologically, and store the first 150 in \texttt{recs}. 

At generation time,  we also currently generate alternative recommendations that include reposts and quote posts from a user's follows; as of November 2025, these recommendations are deployed to a random subset of users. To deploy such alternative recommendations, our feed generator endpoint simply needs to request a different algorithm from the \texttt{recs} database; indeed, the endpoint could request a different algorithm for each user. We also save the first 30 posts from each ranking to a \texttt{counterfactual\_recs} database to facilitate future analyses, so we can see what a user would have viewed if she had checked the feed at that time and if we had used alternative algorithms, unless the user has requested removal from research consideration; this allows us to conduct the type of counterfactual comparison research suggested by \citet{gonzalez2023asymmetric}. 

Finally, since the feed generator endpoint only interfaces with the \texttt{recs} database, we can also split the recommendation generation into independent submodules for different algorithms.  We are currently developing a prototype of an algorithmic feed; when it is deployed we could host this logic in the existing recommendation generation module, or in an independent module that also writes to the \texttt{recs} database.

\paragraph{Cloud Hosting Details} Originally, the generation and dispatch modules were each hosted on separate AWS Lambda functions. The dispatch module was triggered by an AWS EventBridge schedule, and sent generation messages to the generation module using Amazon SQS. Now, generation and dispatch are both Python modules on an AWS EC2 instance, as this scales more efficiently with the number of counterfactual recommendation algorithms. The generation logic is also deployed to AWS Lambda to facilitate the asynchronous refresh that the feed generator triggers when a user accesses the feed. Both \texttt{recs} and \texttt{counterfactual\_recs} are AWS DynamoDB databases.

\paragraph{Cold Start and Onboarding} If the user has checked the feed fewer than ten times at the time of recommendation generation, we prepend a post with details about \skygest, (along with potentially the thread with consent information). 

When a user's Skygest feed contains fewer than 10 posts (because the accounts they follow have not posted enough research content), we append a post encouraging the user to follow more accounts,
and then a ``default'' feed, generated as follows. During the dispatch stage, we also randomly select one of the batches for regenerating the default feed. When we process that batch of recommendations in the recommendations module, we save the most recent 150 posts across all of the posts recommended to users in that batch.  We also show this default feed to logged-out users, since we do not have followers for such users.

\section{Applications of AT Protocol and Paper Skygest}
\label{sec:vision}
Custom feeds on Bluesky are a new avenue for academics to build, deploy, and evaluate social media algorithms with organic users, without requiring a partnership with the platform itself. Our experience with \skygest demonstrates that academics can use custom feeds to host recommenders that garner sustained usage from organic users. We believe that custom feeds, and \skygest in particular, enable study of new questions of interest to the community. We further hope that our architecture description, along with provided code, supports others in this analysis.

\paragraph{Causally Studying the Effects of Algorithm Design}

While the results we present above are observations of user behavior, the infrastructure we built for \skygest is designed to facilitate experimentation (randomized controlled trials). We support using different algorithms for different users, including randomizing across users. We also save counterfactual recommendations for each user across multiple different recommendation algorithms, and collect public interaction data like user likes and reposts over time. We can also insert posts soliciting survey responses in the same way we inserted posts with consent information at the beginning of the feed. With these ingredients, we can compare user engagement and satisfaction across feed designs. (We note that comparing feed designs at the post level will be challenging, due to ranking order effects, cf. \citet{schein2021assessing}).

\paragraph{Designing for User Agency} While custom feeds provide users algorithmic choice through selecting from a pre-defined set of algorithms, recent work has proposed and explored platform designs where users have fine-grained control over their recommendations \citep{betterfeedsknightgeorgetown}, even \textit{within} a feed \citep{jannachcontrol, jahanbakhsh2025value}. For example, \citet{carroll2025ctrlrec} introduced a system to steer recommendations through natural language, Instagram launched a page for users to select topics to include in their feeds \citep{instagram-control}, and \citet{malki2025bonsai} built an interface with a wide range of controls. Bluesky provides an exciting opportunity to deploy controllable feeds in particular because Bluesky users, already familiar with algorithmic choice, may be more interested in organically adopting controllable feeds. One avenue is building an external interface where users can configure their preferences for a custom feed like \skygest (e.g. tuning parameters such as the weight on reposts or content diversity), which are then used to populate the custom feed for that user. \citet{malki2025bonsai} make progress by piloting a user control interface for Bluesky posts, but do not yet integrate it with a deployed custom feed.

\paragraph{Feed Opportunities Beyond \skygest}

Beyond \skygest, a wide variety of interesting open questions exist. For example, feeds with polarizing content could be used to explore novel \textit{bridging} algorithms \citep{wojcik2022birdwatchcrowdwisdombridging}. Feeds focused on entertainment could be used to study overconsumption and related interventions -- e.g., one could explore the effect of a delay or a wellness reminder before a user is shown the feed, as in \citet{onesec}. More broadly -- since the custom feed design does not restrict which posts can be shown, and all Bluesky posts can be ingested via the firehose -- the space of algorithm design is large, in contrast to a commonly faced challenge in browser-extension based re-ranking \citep{piccardi2024reranking}. 

\paragraph{Limitations}

One limitation of custom feed experiments is that these experiments may have low external validity due to idiosyncracies in Bluesky's or the feed's user population. Moreover, custom feed experiments are still subject to several challenges inherent to studying causal effects on social media platforms, such as interference and symbiosis bias \citep{basseinterference,brennansymbiosis2025}. 
Finally, while decentralized social media platforms are an exciting opportunity for academics to build natively on-platform, they still need to acquire organic users by building products users want. 

\section{Conclusion}
\label{sec:conclusion}

In this work, we present \skygest, a custom feed on Bluesky showing personalized posts about papers for academic users. We show quantitative evidence of sustained usage and positive qualitative feedback for \skygest, and use data from \skygest to explore academic Bluesky usage. We provide a detailed overview of our infrastructure to allow others to learn from our deployment experience, and share our full code. Finally, we highlight the potential of custom feeds such as \skygest for facilitating academic research on recommendation algorithms using organic users and outside of industry partnerships.

\section*{Acknowledgements}
The authors would like to thank Cornell's Artificial Intelligence, Practice, and Policy group, and other seminar and group meeting participants. We also thank Rebekah Greenwood for designing the \skygest logo, Adi Tsach for curating a list of \arxiv bot accounts on Bluesky, Devin Gaffney and Joseph Schafer for helpful conversations, and Andr\'{e}s Monroy-Hern\'{a}ndez, Martin Saveski, Mor Naaman, and Robin Burke for pointing us to relevant related work. SG is supported by an NSERC PGS-D fellowship [587665].  NG's work is supported by NSF CAREER IIS-2339427, the Sloan Foundation, NASA, Cornell Tech Urban Tech Hub, Google, Meta, and Amazon research awards. \skygest is built on AWS, supported by AWS credits provided by an Amazon Research Award.

\bibliographystyle{plainnat}
\bibliography{skygest}

\newpage
\appendix

\section{Additional plots of engagement as a function of feed ranking}\label{app:funnel}

\begin{figure}[h]
    \centering
    \begin{subfigure}{0.6\linewidth}
        \includegraphics[width=\linewidth]{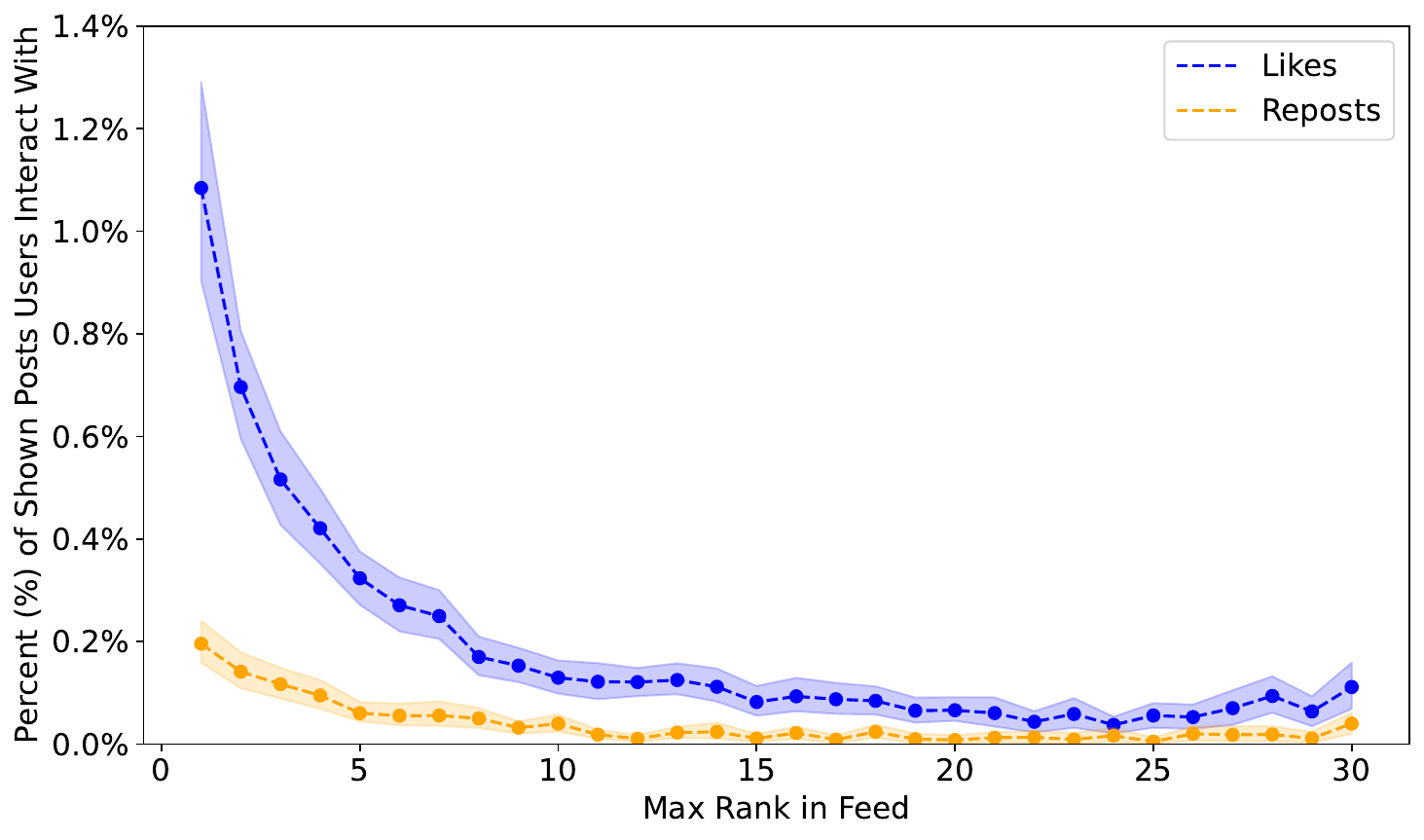}
        \caption{Likes and reposts filtered to within 60 seconds of accessing \skygest, resulting in a total of 8,637 likes and 1,847 reposts.}
    \end{subfigure}

        \begin{subfigure}{0.6\linewidth}
        \includegraphics[width=\linewidth]{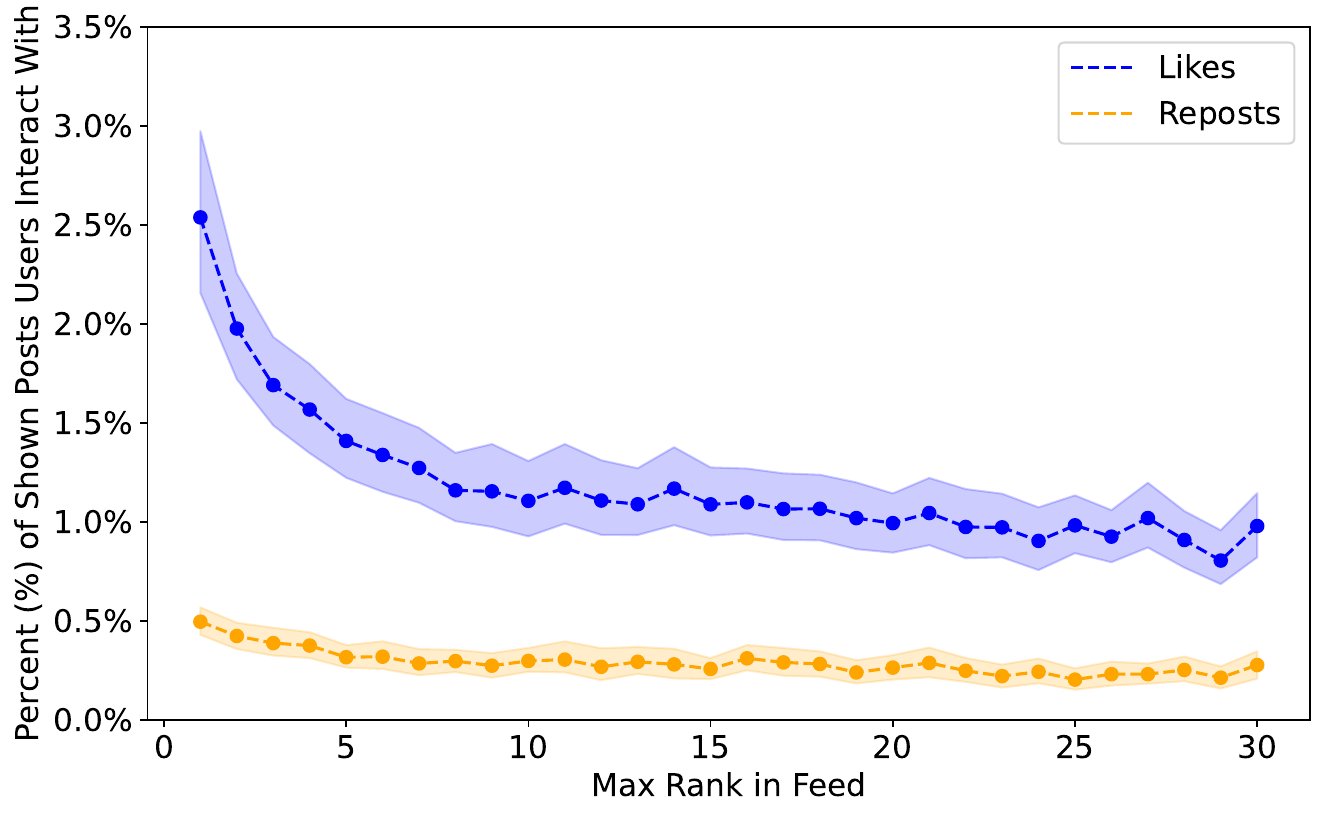}
        \caption{All likes and reposts (no time filter), resulting in a total of 161,669 likes and 41,138 reposts.}
    \end{subfigure}
    \caption{The rate at which users like or repost a post, as a function of how high it appeared in their feed when they switched to it (since users may access the feed multiple times with overlap in posts, we use the highest position that the post appeared). There are strong feed positioning effects. We consider only posts, likes, reposts, and accesses up to September 16, 2025. Bluesky typically requests pages of size 10 and 30; we use access data for cases where Bluesky requests 30 posts. We take the average across post-user pairs for each rank, and report clustered $95\%$ confidence intervals.}
    \label{fig:funnel:additional}
\end{figure}

Figure \ref{fig:funnel} showed the rate at which users interact with a post, as a function of the highest ranking the post appeared in their feed when they viewed \skygest. In Figure \ref{fig:funnel}, we filtered to only interactions which occurred within 30 seconds of accessing \skygest. In Figure \ref{fig:funnel:additional} we repeat this analysis with alternative time windows.

\end{document}